\documentclass[a4paper, oneside, twocolumn, notitlepage, 10pt]{extarticle_ecoc}
\usepackage{ecoc}
\usepackage{url}
\usepackage{amsmath}
\usepackage{amssymb}
\usepackage{amsthm}
\usepackage{authblk}
\usepackage[ruled,vlined]{algorithm2e}
\usepackage{fancyhdr}
\fancypagestyle{first}{%
\chead{\small This paper has been accepted for presentation at the European Conference of Optical Communications (ECOC) 2018 in Rome, Italy. The link to the final version in the Proceedings of ECOC will be posted here once it is available.}
\rhead{\thepage}
\fancyfoot{}
}

\fancypagestyle{others}{%
	\rhead{\thepage}
	\fancyfoot{}
}

\DeclareMathAlphabet{\mathbit}{OML}{cmr}{bx}{it}
\DeclareMathAlphabet{\mathsf}{OT1}{cmss}{m}{n}
\DeclareMathAlphabet{\mathbsf}{OT1}{cmss}{bx}{it}
\newcommand{\inC}[1]{\ensuremath{\in\mathbb{C}^{#1}}}

\newcommand{\inset}[2]{\ensuremath{\in \left\{#1,\ldots,#2\right\}}}
\newcommand{\Real}[1]{\ensuremath{\Re\left\{#1\right\}}}
\newcommand{\Imag}[1]{\ensuremath{\Im\left\{#1\right\}}}

\newlength{\figurewidth}
\newlength{\figureheight}

\usepackage{tikz}
\usetikzlibrary{arrows,decorations,backgrounds,shadows,plotmarks,calc, positioning}
\usetikzlibrary{arrows.meta}

\usepackage{pgfplots}
\pgfplotsset{compat=newest}
\pgfplotsset{plot coordinates/math parser=false}
\pgfplotsset{every axis/.append style={font=\footnotesize}}
\pgfplotsset{
	ylabel right/.style={
		after end axis/.append code={
			\node [rotate=90, anchor=north] at (rel axis cs:1,0.5) {#1};
		}   
	}
}

\begin{document}
\selectlanguage{american}    


\title{Statistics of the Eigenvalues of a Noisy Multi-Soliton Pulse}%


\author{
    Javier Garc\'ia\textsuperscript{(1)}, Vahid Aref\textsuperscript{(2)}
}

\maketitle                  


\begin{strip}
 \begin{author_descr}

   \textsuperscript{(1)} Institute for Communications Engineering, Technische Universit\"at M\"unchen,
   \textcolor{blue}{\underline{javier.garcia@tum.de}}

   \textsuperscript{(2)} Nokia Bell Labs, Stuttgart, Germany,
   \textcolor{blue}{\underline{vahid.aref@nokia-bell-labs.com}}

 \end{author_descr}
\end{strip}

\setstretch{1.1}



\begin{strip}
  \begin{ecoc_abstract}%
For Nonlinear-Frequency Division-Multiplexed (NFDM) systems, the statistics of the received nonlinear spectrum in the presence of additive white Gaussian noise (AWGN) is an open problem. We present a novel method, based on the Fourier collocation algorithm, to compute these statistics.%
  \end{ecoc_abstract}
\end{strip}%
%
%
\thispagestyle{first}
\pagestyle{others}
\section{Introduction}
The Nonlinear Fourier Transform (NFT) has been proposed as an alternative for system design in an attempt to overcome the capacity peak of linear transmission systems over the Nonlinear Schr\"odinger Equation (NLSE) channel~\cite{mansoor_part1}. As a special case, communication using multi-solitons has been demonstrated numerically and experimentally in several scenarios~\cite{turitsyn2017nonlinear}. In the presence of amplifier noise, the eigenvalues and the nonlinear spectrum become strongly correlated. Neglecting this correlation causes a significant loss in the achievable data rates~\cite{buelow2018MI,gui2017alternative}.
Therefore, knowing the statistics of the nonlinear spectrum is crucial to optimally design and detect multi-soliton pulses.

The effect of channel noise on the nonlinear spectrum of a pulse is not yet well understood. The analytic result is available only for few special cases, such as a first-order soliton~\cite{zhang_perturbation,derevyanko_pdf}. 
Wahls~\cite{sander_statistics} proposed a numerical method to compute the statistics of the spectral coefficients when a signal is contaminated by AWGN. 
Although applicable for any sampled pulse, this method depends on the time-domain representation of the pulse, not directly its nonlinear spectrum, and has the same NFT algorithmic challenges in numerical accuracy and complexity. 

We present a novel method, developed in frequency domain, to compute the 
statistics of the eigenvalues of a 
multi-soliton contaminated by AWGN. 
We show that they have a jointly complex Gaussian distribution with a covariance matrix formulated in terms of the nonlinear spectrum.  We show the accuracy of our method for a $2$-soliton through simulations.
Although we develop our method for multi-solitons, its application can be extended for other NFDM pulses.

\section{Nonlinear Fourier Transform and Multi-Solitons} 
The NFT of a signal $q(t)$ is
obtained by solving the Zakharov-Shabat system (ZSS):
\begin{equation}
\left(\begin{matrix}
-\frac{\partial}{\partial t} & q(t) \\ q^*(t) & \frac{\partial}{\partial t}
\end{matrix}\right)\left(\begin{matrix}
v_1(t,\lambda) \\ v_2(t,\lambda)
\end{matrix}\right)=j\lambda \left(\begin{matrix}
v_1(t,\lambda) \\ v_2(t,\lambda)
\end{matrix}\right)
\label{eq:zs_0}
\end{equation}
\begin{equation}
v(t, \lambda)\to\left(\begin{matrix}
1\\0
\end{matrix}\right)e^{-j\lambda t}, \quad t\to-\infty
\end{equation}
where $\lambda$ is the eigenvalue and the \textit{Jost solution} $v(t,\lambda)=(v_1\ v_2)^T$ is the eigenvector. 
The \textit{spectral coefficients} $a(\lambda)$ and $b(\lambda)$ are given by
	\begin{align*}
	a(\lambda)&=\lim\limits_{t\to+\infty} v_1e^{j\lambda t} &
	b(\lambda)&=\lim\limits_{t\to+\infty} v_2e^{-j\lambda t}.
	\end{align*}
The nonlinear spectrum consists of two parts:\\
$(i)$ \textit{continuous spectrum}: $Q_c(\lambda)=\frac{b(\lambda)}{a(\lambda)}$, for $\lambda\in\mathbb{R}$,\\
$(ii)$ \textit{discrete spectrum}: $Q_d(\lambda_k)=\frac{b(\lambda_k)}{a_\lambda(\lambda_k)}$, for the $K$ distinct eigenvalues $\left\{\lambda_k\in\mathbb{C}^+\colon a(\lambda_k)=0\right\}$ \\
where $a_{\lambda}=\mathrm{d}a/\mathrm{d}\lambda$. The nonlinear spectrum of a signal $q(z, t)$ propagating along a noiseless fiber (modeled by the NLSE) evolves in $z$ according to:
\begin{subequations}
	\begin{align}
	Q_c(z, \lambda)&=Q_c(0, \lambda)e^{4j\lambda^2 z} \label{eq:Q_cont}\\
	\lambda_k(z)&=\lambda_k(0) \label{eq:lambda_z}\\
	Q_d(z, \lambda_k)&=Q_d(0, \lambda_k)e^{4j\lambda_k^2 z}. \label{eq:Qd_z}
	\end{align}
\end{subequations}
\textit{Multi-soliton} pulses are a special class of pulses with no continuous spectrum, i.e., $Q_c(\lambda)=0$. From~\eqref{eq:lambda_z}, the eigenvalues of a multi-soliton stay constant along propagation, and the spectral amplitudes evolve independently of each other. 
The complete version of the Darboux algorithm~\cite{aref_control_detection} (Algorithm ~\ref{algo:darboux}) constructs a multi-soliton and its Jost solutions from the discrete spectrum.
\\
\textit{Fourier Collocation (FC) Method}~\cite{yang_nlse}: The FC Method represents the ZSS~\eqref{eq:zs_0} in terms of the Discrete Fourier Transform (DFT) of the sampled pulse $q_m=q(mT_s)$ and the Jost vectors $v_k[m]=(v_{k,1}[m]\ v_{k,2}[m])^T\triangleq v(mT_s, \lambda_k)$, where $m\inset{-N}{N}$ and $T_s$ is the sampling period:
\vspace{-0.1em}
\begin{equation*}
q_m=\sum_{n=-N}^{N}c_ne^{jn\frac{2\pi}{M}m}, \quad v_{k,i}[m]=\sum_{n=-N}^{N}a_{k,i}[n]e^{jn\frac{2\pi}{M}m}
\label{eq:fs}
\end{equation*}
for $i\in\{1,2\}$ and $M=2N+1$. The ZSS~\eqref{eq:zs_0} becomes
\begin{equation}
\mathbf{S}\mathbf{a}_k=\lambda_k\mathbf{a}_k
\label{eq:fc}
\end{equation}
where
\begin{equation}
\mathbf{S}=\left(\begin{matrix}\boldsymbol{\Omega} & \boldsymbol{\Gamma} \\ -\boldsymbol{\Gamma}^H & -\boldsymbol{\Omega} \end{matrix}\right)
\label{eq:T}
\end{equation}
\begin{equation*}
\mathbf{a}_k=\left(
a_{k,1}[-N] \ \cdots \ a_{k,1}[N] \ a_{k,2}[-N] \ \cdots \ a_{k,2}[N]
\right)^T
\label{eq:A1A2}
\end{equation*}
$\boldsymbol{\Omega}=-\frac{2\pi}{L}\mathrm{diag}\left(-N,\ldots,N\right)$, and $\boldsymbol{\Gamma}\inC{M\times M}$ is a Toeplitz matrix whose first column is $-j(c_0\ \dots\ c_N\ 0\ \dots\ 0)^T$ and whose first row is $-j(c_0\ \dots\ c_{-N}\ 0\ \dots\ 0)$. 

As a result, the eigenvalues and the right eigenvectors of $\mathbf{S}$ correspond to the eigenvalues and the DFT of the Jost solutions of \eqref{eq:zs_0}. Although our analysis is in the limit of $N$, FC is known to estimate the eigenvalues with a fine precision for rather small values of $N$. 

\section{Statistics of the Eigenvalues}
Consider a sampled multi-soliton perturbed by AWGN
\begin{equation}
\hat{q}_m=q_m+\tilde{q}_m
\end{equation}
where $\tilde{q}_m$ are i.i.d. circularly symmetric complex Gaussian variables with zero mean and variance $\sigma_{\tilde{q}}^2$. The DFT of $\hat{q}_m$
is $\hat{c}_n=c_n+\tilde{c}_n$, where $\tilde{c}_n$ is also i.i.d. circularly symmetric complex Gaussian with per-sample variance $\sigma_{\tilde{c}}^2=\sigma_{\tilde{q}}^2/M$. In~\eqref{eq:fc}, $\mathbf{S}$ is replaced by the perturbed $\hat{\mathbf{S}}=\mathbf{S}+\tilde{\mathbf{S}}$, with 
\begin{equation}
\tilde{\mathbf{S}}=\left(\begin{matrix}
\mathbf{0} & \tilde{\boldsymbol{\Gamma}} \\
-\tilde{\boldsymbol{\Gamma}}^H & \mathbf{0}
\end{matrix}\right)
\label{eq:T_prime}
\end{equation}
where $\tilde{\boldsymbol{\Gamma}}$ is defined similarly to $\boldsymbol{\Gamma}$ with all $c_n$ replaced by $\tilde{c}_n$. Using perturbation theory~\cite{kato_eigenvalues},
we can show that $\hat{\mathbf{S}}$ has eigenvalues
\begin{equation}\label{eq:lambda_pert}
\hat{\lambda}_k=\lambda_k+\tilde{\lambda}_k+o\left(1/\mathrm{SNR}\right)
\end{equation}
if $\mathrm{SNR}\triangleq\sum_n \left|c_n\right|^2/(M\sigma_{\tilde{c}}^2)\gg 1$. The term $\tilde{\lambda_k}$ is
\begin{equation}
\label{eq:eigenvalue_perturb}
\tilde{\lambda}_k= \frac{\left(\mathbf{b}_{k,1}\tilde{\boldsymbol{\Gamma}}\mathbf{a}_{k,2}-\mathbf{b}_{k,2}\tilde{\boldsymbol{\Gamma}}^H\mathbf{a}_{k,1}\right)}{\mathbf{b}_{k,1}\mathbf{a}_{k,1}+\mathbf{b}_{k,2}\mathbf{a}_{k,2}}
\end{equation}
where $\mathbf{a}_{k,1}$ and $\mathbf{a}_{k,2}$ are respectively the first and second halves of $\mathbf{a}_k$, and $\mathbf{b}_{k,1}$ and $\mathbf{b}_{k,2}$ are row vectors with $\mathbf{b}_{k,1}[m]=\mathbf{a}_{k,2}[-m]$ and $\mathbf{b}_{k,2}=\mathbf{a}_{k,1}[-m]$, for $-N\leq m\leq N$.

\begin{algorithm}[t]
	\BlankLine
		\For{$i\leftarrow 1$ \KwTo $K$}{
			$b(\lambda_i)=\frac{Q_d(\lambda_i)}{\lambda_i-\lambda_i^*}\prod_{k=1,k\ne i}^K \frac{\lambda_i-\lambda_k}{\lambda_i-\lambda_k^*}$\;
			$v^{(0)}_i(t)=(e^{-j\lambda_i t},-b(\lambda_i) e^{j\lambda_i t})^T$\;
		}
		$q^{(0)} = 0$\;
		\tcc{iteratively add $(\lambda_i,Q_d(\lambda_i))$}
		\For{$i\leftarrow 1$ \KwTo $K$}{
			$(\psi_1,\psi_2)= v_i^{(i-1)}(t)$\;
			\tcc{update signal}
			$q^{(i)}(t)= q^{(i-1)}(t)-2j(\lambda_i-\lambda_i^*)\frac{\psi_2^*(t)\psi_1(t)}{|\psi_1(t)|^2+|\psi_2(t)|^2}$\;
			
			\tcc{update JS for $\lambda_i$}
			$C=b(\lambda_i) \prod_{k=1}^{i-1}\left(\lambda_i-\lambda_k\right) \prod_{k=i+1}^{K}1/\left(\lambda_i-\lambda_k^*\right)$\;
			$\displaystyle v_i^{(i)}(t)=\frac{C}{|\psi_1(t)|^2+|\psi_2(t)|^2}\left(\begin{matrix}-\psi_2^*(t)\\\psi_1^*(t)\end{matrix}\right)$\;
			
			\tcc{update JS for $\lambda_{k\ne i}$}
			\For{$k\leftarrow 1$ \KwTo $K$; $k\ne i$}{
				$v_{k,1}^{(i)}(t)= \left(\lambda_k -\lambda_i^*-\frac{(\lambda_i-\lambda_i^*)|\psi_1(t)|^2}{|\psi_1(t)|^2+|\psi_2(t)|^2}
				\right) v_{k,1}^{(i-1)}(t) -\frac{(\lambda_i-\lambda_i^*)\psi_2^*(t)\psi_1(t)}{|\psi_1(t)|^2+|\psi_2(t)|^2}v_{k,2}^{(i-1)}(t)$\;
				$v_{k,2}^{(i)}(t)= 
				-\frac{(\lambda_i-\lambda_i^*)\psi_2(t)\psi_1^*(t)}{|\psi_1(t)|^2+|\psi_2(t)|^2}v_{k,1}^{(i-1)}(t)
				+\left(\lambda_k -\lambda_i+\frac{(\lambda_i-\lambda_i^*)|\psi_1(t)|^2}{|\psi_1(t)|^2+|\psi_2(t)|^2}
				\right)v_{k,2}^{(i-1)}(t)$\;
			}
			%
		}
		
	\caption{Darboux Transform to compute $K-$soliton $q^{(K)}(t)$ and Jost Solutions (JS) $v_k^{(K)}(t), k=1,\dots,K$.}
	\label{algo:darboux}
\end{algorithm}

\begin{figure*}[t]\centering
	\setlength{\figurewidth}{0.27\linewidth}
	\setlength{\figureheight}{0.618\figurewidth}
%
%
\definecolor{mycolor1}{rgb}{0.00000,0.44700,0.74100}%
\definecolor{mycolor2}{rgb}{0.85000,0.32500,0.09800}%
\begin{tikzpicture}

\begin{axis}[%
name=ax1,
width=0.951\figurewidth,
height=\figureheight,
at={(0\figurewidth,0\figureheight)},
scale only axis,
xmin=0,
xmax=1,
xlabel style={font=\color{white!15!black}},
xlabel={Phase difference ($\times 2\pi$)},
ymin=0,
ymax=0.099,
ylabel style={font=\color{white!15!black}},
ylabel={NMSE},
axis background/.style={fill=white},
title style={font=\bfseries},
legend style={at={(0.03,0.97)}, anchor=north west, legend cell align=left, align=left, draw=white!15!black}
]
\addplot [color=mycolor1, thick, dashed]
table[row sep=crcr]{%
	0	0.0438929412055956\\
	0	0.0342974428167118\\
	0	0.0334737376195604\\
	0.015625	0.0408337085019224\\
	0.03125	0.0377936945829512\\
	0.046875	0.0370404235427028\\
	0.0625	0.0254088417641027\\
	0.078125	0.0390804952579099\\
	0.09375	0.026217021183972\\
	0.109375	0.0373542546349976\\
	0.125	0.0312497625777969\\
	0.140625	0.0274563008047645\\
	0.15625	0.0242915924172094\\
	0.171875	0.028403401808997\\
	0.1875	0.0268228387833505\\
	0.203125	0.0267588863643901\\
	0.21875	0.0259708047473425\\
	0.234375	0.0325370654498849\\
	0.25	0.0279599768308911\\
	0.265625	0.022794447472779\\
	0.28125	0.0221579602496013\\
	0.296875	0.021520166158123\\
	0.3125	0.0168275474099406\\
	0.328125	0.0199297114702561\\
	0.34375	0.0131158380194666\\
	0.359375	0.0120784976537797\\
	0.375	0.00756579240080744\\
	0.390625	0.00899674431389565\\
	0.40625	0.00880890267091259\\
	0.421875	0.0226074812137722\\
	0.4375	0.0318595508754612\\
	0.453125	0.0360759906828743\\
	0.46875	0.0626750132630081\\
	0.484375	0.06138180873454\\
	0.5	0.0805367145918127\\
	0.515625	0.0946898551844713\\
	0.53125	0.0640455551772085\\
	0.546875	0.0451385524212998\\
	0.5625	0.0350670554858777\\
	0.578125	0.0133846898551677\\
	0.59375	0.0153283902154372\\
	0.609375	0.0100858088032758\\
	0.625	0.00971550857434959\\
	0.640625	0.0116454436168104\\
	0.65625	0.0197211758079761\\
	0.671875	0.0233051946123451\\
	0.6875	0.0301951474706691\\
	0.703125	0.0344843747343327\\
	0.71875	0.0410934332499866\\
	0.734375	0.0489258750011816\\
	0.75	0.0582269872688845\\
	0.765625	0.0618577155306327\\
	0.78125	0.0617519138526419\\
	0.796875	0.0653720716501394\\
	0.8125	0.0609509663325561\\
	0.828125	0.0596986905978825\\
	0.84375	0.0570810127749495\\
	0.859375	0.0716117420759658\\
	0.875	0.0537663524283255\\
	0.890625	0.0501532514548613\\
	0.90625	0.0436502489229558\\
	0.921875	0.0518345342283986\\
	0.9375	0.0408107443388283\\
	0.953125	0.0494935293248697\\
	0.96875	0.0445690697448925\\
	0.984375	0.0314754446977629\\
};
\addlegendentry{FC}

\addplot [color=mycolor2, thick]
table[row sep=crcr]{%
	0	0.00724712718942051\\
	0	0.00451789895268626\\
	0	0.00463112626727505\\
	0.015625	0.00716113097140815\\
	0.03125	0.00606879840798556\\
	0.046875	0.00870350061008514\\
	0.0625	0.00457398698978338\\
	0.078125	0.0111355704059567\\
	0.09375	0.00614232201103638\\
	0.109375	0.0117456983525269\\
	0.125	0.0106126776063319\\
	0.140625	0.0113858149166819\\
	0.15625	0.0120846538981416\\
	0.171875	0.0166413993134303\\
	0.1875	0.0185711242097264\\
	0.203125	0.0168356644657599\\
	0.21875	0.0180990148833633\\
	0.234375	0.0217511708241528\\
	0.25	0.0185597940067986\\
	0.265625	0.0178382235690695\\
	0.28125	0.0187102558933383\\
	0.296875	0.0184532217117993\\
	0.3125	0.0152389476257675\\
	0.328125	0.0148716929144198\\
	0.34375	0.00907383338738494\\
	0.359375	0.00807545529604104\\
	0.375	0.00441284587493444\\
	0.390625	0.00319584048177214\\
	0.40625	0.00260333604041952\\
	0.421875	0.00348202703934432\\
	0.4375	0.00326955942896351\\
	0.453125	0.00426012238841849\\
	0.46875	0.00670180963228125\\
	0.484375	0.00944108432841625\\
	0.5	0.0136654608503675\\
	0.515625	0.0118147879937473\\
	0.53125	0.00802674709055377\\
	0.546875	0.00326563972318302\\
	0.5625	0.00393679689286706\\
	0.578125	0.00170662525771033\\
	0.59375	0.00440960955744104\\
	0.609375	0.00571313064708719\\
	0.625	0.00681409185267513\\
	0.640625	0.00802209499436866\\
	0.65625	0.0146278162910399\\
	0.671875	0.0147806173632247\\
	0.6875	0.018555670214636\\
	0.703125	0.0219293905692395\\
	0.71875	0.0248392606073682\\
	0.734375	0.0260347003767754\\
	0.75	0.0321702238112612\\
	0.765625	0.0300105706554648\\
	0.78125	0.0284989301585723\\
	0.796875	0.028742864300828\\
	0.8125	0.0252577505405501\\
	0.828125	0.0238768696589349\\
	0.84375	0.021879337645522\\
	0.859375	0.0271587539590636\\
	0.875	0.0161992147310861\\
	0.890625	0.0155092800895491\\
	0.90625	0.0109756170671631\\
	0.921875	0.0132839470512089\\
	0.9375	0.00997383113416576\\
	0.953125	0.0118028923482873\\
	0.96875	0.00821576758641634\\
	0.984375	0.0051797008557006\\
};
\addlegendentry{NS}
\end{axis}
\node at (ax1.north) [inner sep=0.05em, anchor=south] {\footnotesize(a)};
\end{tikzpicture}%
	\input{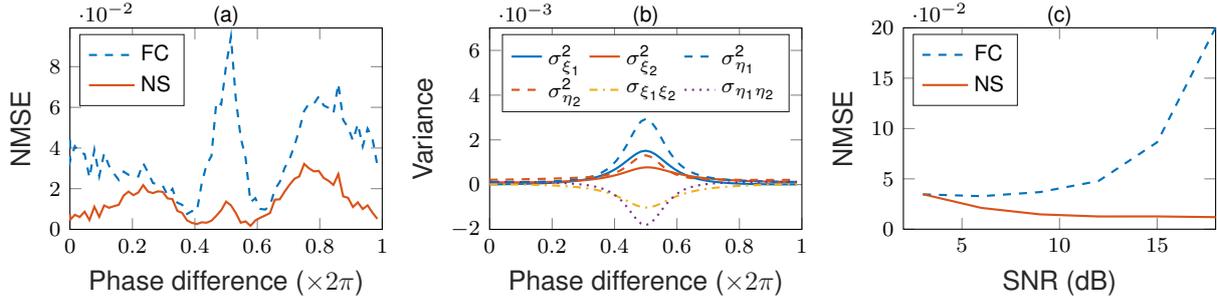}
%
%
\definecolor{mycolor1}{rgb}{0.00000,0.44700,0.74100}%
\definecolor{mycolor2}{rgb}{0.85000,0.32500,0.09800}%
\begin{tikzpicture}

\begin{axis}[%
name=ax3,
width=0.951\figurewidth,
height=\figureheight,
at={(0\figurewidth,0\figureheight)},
scale only axis,
xmin=2,
xmax=18,
xlabel style={font=\color{white!15!black}},
xlabel={SNR (dB)},
ymin=0,
ymax=0.2,
scaled y ticks={base 10:2},
ylabel style={font=\color{white!15!black}},
ylabel={NMSE},
axis background/.style={fill=white},
legend style={at={(0.03, 0.97)}, anchor=north west, legend cell align=left, align=left, draw=white!15!black}
]
\addplot [color=mycolor1, dashed, thick]
  table[row sep=crcr]{%
3	0.0346256822739236\\
6	0.032797987248807\\
9	0.0368735696397518\\
12	0.0478774131495377\\
15	0.0863153521600286\\
18	0.200302791960325\\
};
\addlegendentry{FC}

\addplot [color=mycolor2, thick]
  table[row sep=crcr]{%
3	0.0347491924591135\\
6	0.0211345060069872\\
9	0.014739521899742\\
12	0.0126518528427625\\
15	0.012675868109553\\
18	0.0119727681147205\\
};
\addlegendentry{NS}

\end{axis}
\node at (ax3.north) [inner sep=0.05em, anchor=south] {\footnotesize(c)};
\end{tikzpicture}%
	\caption{(a) NMSE of eigenvalue covariance matrix estimation for a $2$-soliton with $256$ samples and SNR$=10$ dB. (b) Entries of the covariance matrix for the same experiment. (c) $\phi$-averaged NMSE of eigenvalue covariance matrix estimation for a $2$-soliton with $256$ samples. FC: Fourier collocation. NS: Newton search.}
	\label{fig:nmse_var}
\end{figure*}

Let $\tilde{\lambda}_k=\tilde{\xi}_k+j\tilde{\eta}_k$ and $\tilde{\boldsymbol{\xi}}=(\tilde{\xi}_1,\dots,\tilde{\xi}_K)^T$ and $\tilde{\boldsymbol{\eta}}=(\tilde{\eta}_1,\dots,\tilde{\eta}_K)^T$. Reordering \eqref{eq:eigenvalue_perturb},
we have 
\begin{equation}
\left(\begin{matrix}\tilde{\boldsymbol{\xi}} \\ \tilde{\boldsymbol{\eta}}\end{matrix}\right)=\mathbf{D}\left(\begin{matrix}
\Real{\tilde{\mathbf{c}}} \\ \Imag{\tilde{\mathbf{c}}}
\end{matrix}\right)
\label{eq:pert_ri}
\end{equation}
where $\tilde{\mathbf{c}}=\left(\tilde{c}_{-N} \ \cdots \ \tilde{c}_{N}\right)^T$ and, for $k\inset{1}{K}$, the rows of $\mathbf{D}$ are
\begin{align}
\mathbf{D}_{k,\colon}&=\left[\begin{matrix}
\Imag{\overline{\mathbf{r}}_{k,2}+\mathbf{r}_{k,1}} & \Real{\overline{\mathbf{r}}_{k,2}-\mathbf{r}_{k,1}}
\end{matrix}\right] \nonumber\\
\mathbf{D}_{k+K,\colon}&=\left[\begin{matrix}
\Real{-\overline{\mathbf{r}}_{k,2}-\mathbf{r}_{k,1}} & \Imag{\overline{\mathbf{r}}_{k,2}-\mathbf{r}_{k,1}}
\end{matrix}\right] \nonumber
\end{align}
where we define for $n\inset{-N}{N}$:
\begin{equation*}
\label{eq:autocor}
r_{k, i}[n]=\frac{1}{g_k}\sum_{p=-N}^{N}a_{k,i}[p]a_{k,i}[n-p],\quad i\in\left\{1,2\right\}
\end{equation*}
and $\overline{\mathbf{r}}_{k,2}[n]=\mathbf{r}_{k,2}[-n]$. Moreover, $g_k=2\sum_{n=-N}^N a_{k,1}[ n]a_{k,2}[ -n]$. 

Since $\tilde{c}_n$ are i.i.d. Gaussian, $(\tilde{\boldsymbol{\xi}}^T,\tilde{\boldsymbol{\eta}}^T)^T$ is jointly Gaussian with zero mean and covariance matrix
\begin{equation}
\label{eq:covariance}
\mathbf{C}_{\xi\eta,\xi\eta}=\frac{\sigma_{\tilde{q}}^2}{2M}\mathbf{D}\mathbf{D}^H.
\end{equation}
Note that Yousefi~\cite{mansoor_part3} also uses eigenvalue perturbation theory to obtain a result similar to~\eqref{eq:lambda_pert}. Our contribution is a procedure (Algorithm~\ref{algo:darboux}) to numerically compute the Jost solutions required in~\eqref{eq:lambda_pert}, and the computation of the full joint probability distribution of the eigenvalues in the frequency domain. 

\section{Numerical validation}\label{sec:numerical}

We ran a numerical experiment with a $2$-soliton with $Q_d(0.3j)=1.8$ and $Q_d(0.6j)=3.6e^{j\phi}$, for varying $\phi$. %
The time axis had $256$ samples in $[-12.4842, 12.4842]$. The per-sample variance of the AWGN was $\sigma_{q'}^2=0.014362$ ($\mathrm{SNR}=10\;\mathrm{dB}$).

For $\phi\in\left[0, 2\pi\right]$, the analytic covariance matrix $\mathbf{C}_{\xi\eta,\xi\eta}$ from~\eqref{eq:covariance} was compared with experimental covariance matrices from both the Fourier collocation (FC) method and the Newton-Raphson search (NS) method with forward-backward computation and trapezoidal rule~\cite{aref_control_detection}. The number of Monte-Carlo runs for each value of $\phi$ was $12288$. Fig.~\ref{fig:nmse_var} (a) shows the normalized mean square error (NMSE) of the estimation of $\mathbf{C}_{\xi\eta,\xi\eta}$:
\begin{equation}
\mathrm{NMSE}=\frac{\left\|\hat{\mathbf{C}}_{\xi\eta,\xi\eta}-\mathbf{C}_{\xi\eta,\xi\eta}\right\|_F^2}{\left\|\mathbf{C}_{\xi\eta,\xi\eta}\right\|_F^2}
\end{equation}
where $\hat{\mathbf{C}}_{\xi\eta,\xi\eta}$ is the experimental covariance matrix. Newton search is more robust against noise. However, observe that our analytic formula~\eqref{eq:covariance} matches the experimental results very well ($\mathrm{NMSE}\approx 0.01$). FC seems to be inaccurate when $\phi$ is around $\pi$. Note that the NMSE is a combination of the errors from [a] the NFT algorithm (NS or FC), [b] limited amount of Monte-Carlo runs, and [c] neglecting higher-order perturbation terms. Simulations with higher sampling rate and more Monte-Carlo runs showed that [a] and [b] play a negligible role in Fig.~\ref{fig:nmse_var}.

Fig.~\ref{fig:nmse_var} (b) plots the relevant entries of $\mathbf{C}_{\xi\eta,\xi\eta}$. The variances of the imaginary parts $\eta$ of the eigenvalues (dashed lines) are larger than those of the real parts $\xi$ (solid lines), and there is correlation between the two eigenvalues. Solitons with $\phi=\pi$ are less robust to noise.


In a second experiment, we varied the SNR and obtained the average NMSE over different phase differences $\phi$. The results are plotted in Fig.~\ref{fig:nmse_var} (c). Again,~\eqref{eq:covariance} accurately predicts the covariance matrix of the eigenvalues obtained using Newton search. The covariances become smaller at high SNR, and the increase of NMSE with SNR for the FC method hints at numerical instability.

\section{Conclusion}\label{sec:conclusion}
We proposed a new frequency-domain method based on Fourier collocation to compute the statistics of the eigenvalues of a multi-soliton perturbed by AWGN. The method is shown to match the simulation results for a $2$-soliton.


\bibliographystyle{ecoc_bib}
\bibliography{ecoc-2018}
\vspace{-4mm}

\end{document}